\documentclass[prl,amssymb,amsfonts,amsmath,showpacs,twocolumn]{revtex4}
\usepackage[dvips]{graphicx}
\newcommand{\pe}{\mbox{\footnotesize pe}}
\newcommand{\tst}{\mbox{\footnotesize sat}}
\newcommand{\tmn}{\mbox{\footnotesize min}}
\newcommand{\tht}{\mbox{\footnotesize th}}
\newcommand{\tr}{\mbox{\footnotesize tr}}

\begin{document}

\title{Nontrapping arrest of Langmuir wave damping near
the threshold amplitude}

\author{A.~V. Ivanov}
\author{Iver~H. Cairns}

\affiliation{School of Physics, The University of Sydney, NSW
2006, Sydney, Australia}

\begin{abstract}

Evolution of a Langmuir wave is studied numerically for finite
amplitudes slightly above the threshold which separates damping from
nondamping cases. Arrest of linear damping is found to be a
second-order effect due to ballistic evolution of perturbations,
resonant power transfer between field and particles, and organization
of phase space into a positive slope for the average distribution
function $f_{av}$ around the resonant wave phase speed $v_\phi$. Near
the threshold trapping in the wave potential does not arrest damping
or saturate the subsequent growth phase.

\end{abstract}


\maketitle

\begin{figure}[b]
\includegraphics[clip]{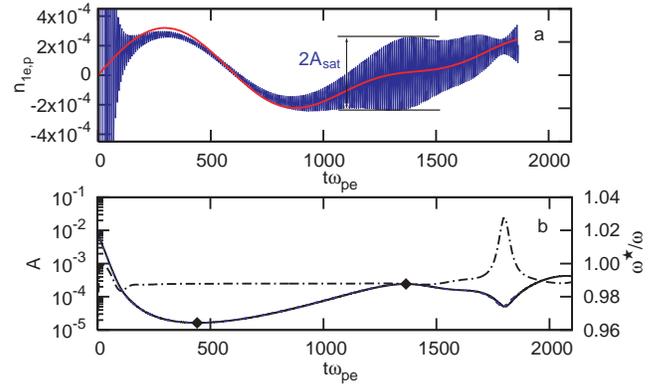}
\caption{(a) Two-component electron (in blue) and ion (red line)
$m=1$ field evolution for $m_p/m_e=1836$. (b) Electric field envelope
for mode $m=1$ for immobile ions (black solid line) and mobile ions
(blue dashed line), and the ratio of the simulated frequency to its
analytic prediction, $\omega^\star / \omega$ (right axis, black
dash-dotted line). Diamonds mark the ``arrest'' time $t_{\tmn}$ and
``saturation'' time $t_{\tst}$.\label{f1}}
\end{figure}

Plasma theory has usually been pursued independently of the theory
of critical phenomena. Recently, however, it has been revealed
that evolution of a monochromatic electrostatic Langmuir wave of
finite amplitude in a Maxwellian plasma is a threshold phenomenon.
Specifically, after a short initial period of approximately linear
damping according to Landau's \cite{landau} classic theory, a wave
with initial amplitude $A_0$ greater than a threshold
$A^{\star}_0$ stops decreasing and starts to grow approximately
exponentially before undergoing irregular oscillations in
amplitude \cite{sug-kam,iv-ca-ro} (Fig.~\ref{f1}). Both the
amplitudes and times at which the wave first ceases to damp and
grow (labelled ``arrest'' and ``saturation'') are power-law
functions of the difference $(A_0 - A^{\star}_0)$ \cite{iv-ca-ro},
thus casting the process into a \emph{new} universality class of
dynamic critical phenomena.

For a collisionless plasma the distribution function (DF) is
usually not Gaussian, and because of the long-range character of
the Coulomb force these systems are outside the domain of
equilibrium thermodynamics. Unlike the theory of critical
phenomena in thermodynamics, where only \emph{spatial}
correlations are considered through the order parameter $\phi({\bf
x})$ and the partition function $Z = \int \mathcal{D}\phi {\bf
(x)} e^{-H[\phi]}$ \cite{goldn_hinr}, threshold physics in
collisionless systems involves correlations in \emph{velocity}
space \cite{ivanov}. Therefore these correlations, due to resonant
energy exchange between particles and waves, are a new paradigm
for critical phenomena potentially applicable in a vast class of
systems, e.g. coupled phase oscillators which show Landau damping
or equivalent Josephson-junction arrays \cite{strogatz}.

Crawford's pioneering analysis \cite{crawford} reveals the
striking difference between thermodynamic and plasma situations
due to this physics: the resonance between particles and waves at
the phase velocity $v_\phi=\omega_{\pe}/k$ turns the thermodynamic
exponent $\beta=1/2$ \cite{goldn_hinr} into the ``trapping
scaling" exponent $\beta=2$, which describes saturation of the
weak bump-on-tail \cite{bump} and gravitational instabilities
\cite{ivanov}. In the frameworks of linear and quasilinear theory
\cite{qlt} arrest of the linear damping of plasma waves (as well
as saturation of the growth) might be explained in terms of
flattening of the DF at $v_\phi$, thus bringing the damping
(growth) rate $\gamma_{\mbox{\scriptsize L}} \sim (\partial
f/\partial v)_{v=v_{\phi}}$ of a kinetic instability to zero.

Trapping of electrons in a monochromatic wave's electric potential is
often suggested as a nonlinear mechanism to stop the initial
exponential damping phase and to saturate the wave's growth
\cite{oneil,la_do}. Trapping and its associated
Bernstein-Greene-Kruskal (BGK) modes \cite{bgk} also imply a certain
shape of the DF plus trapped and untrapped orbits in velocity phase
space. However, it is controversial whether trapping is relevant to
the damping threshold. For instance, one analysis \cite{firpo}
assumes ergodicity of trapped particles in a single-wave potential
and predicts the threshold initial electric field amplitude $E_0$
through the critical ratio $q_c = |\gamma_ {\mbox{\scriptsize
L}}|/\omega_{\tr} \approx 0.06$ of the absolute Landau damping rate
$|\gamma_{\mbox{\scriptsize L}}|$ to the trapping frequency
$\omega_{\tr} = (k E_0 e / m_{e})^{1/2}$. In contrast full
Vlasov-Poisson (V-P) simulations for a Maxwellian plasma yield $q_c
\approx 0.85$ from the asymptotic evolution \cite{brun} and $q_c
\approx 1.0$ from the initial evolution \cite{iv-ca-ro}, with
constants of proportionality slightly different from unity for other
thermal plasmas \cite{iv-ca-ro}.

Other conflicting evidence exists on the role of trapping.
Consider the critical exponents $\tau_{\tmn}$, $\beta_{\tmn}$,
$\tau_{\tst}$ and $\beta_{\tst}$ for the power-law functions of
$(A_0 - A^{\star}_0)$ obeyed by, respectively, the time $t_{\tmn}$
and amplitude $A_{\tmn}$ at which the initial damping phase
finishes, as well as the time $t_{\tst}$ and amplitude $A_{\tst}$
at which the first exponential growth phase saturates
\cite{iv-ca-ro}: e.g., $t_{\tmn} \propto (A_0-A^{\star}_0)
^{-\tau_{\mbox{\footnotesize min}}}$ and $A_{\tmn} \propto
(A_0-A^{\star}_0)^{\beta_{\mbox{\footnotesize min}}}$. First, the
temporal exponents $\tau_{\tmn}=0.901 \pm 0.008$ and
$\tau_{\tst}=1.039 \pm 0.011$ are measurably different from each
other and the value $0.5$ expected from the definition of
$\omega_{\mbox{\footnotesize tr}}$. Second, the field exponents
$\beta_{\tst}=1.88 \pm 0.07$ and $\beta_{\tmn}=2.72 \pm 0.09$ are
remarkably different from each other and the value
$\beta_{\mbox{\footnotesize tr}}=1$ expected for trapping
\cite{la_do}. These points argue against trapping
causing either of the arrest and saturation phenomena. Third,
calculations with $A_{0} \gg A^{\star}_0$ lead to $\tau$
and $\beta$ exponents closer to 0.6 and 1.3, respectively, and the
oscillation spectrum has clear peaks near
$\omega_{\mbox{\footnotesize tr}}$, suggesting that trapping plays
a role well above threshold \cite{iv-ca-ro}.

In this Letter we first simulate one-dimensional (1-D) V-P
two-component plasma with initially Maxwellian distributions for
electrons and ions and demonstrate that ion mobility does not
affect the threshold phenomenon for Langmuir wave damping seen in
V-P simulations without ions. Then, using one-component electron
V-P simulations, we demonstrate that the DF phase portrait when
the wave first ceases to damp is much simpler than a BGK
equilibrium \cite{bgk} and shows no evidence for trapping.
Instead, we demonstrate that the initial DF resonantly evolves a
positive slope in velocity space that stops the initial Landau
damping and supports the subsequent exponential growth. We also
demonstrate that the DFs are different at the arrest and
saturation times and are not consistent with trapping.

To clarify the importance of ion mobility we employ first the
two-component 1-D V-P model, normalizing to electron quantities:
\begin{eqnarray}
\label{dst}
     && \partial f_a/\partial t + v \, \partial f_a/\partial x
    - \mu_{a} E \, \partial f_a/\partial v = 0 ~, \\
\label{pot}
    && \partial E/\partial x = \int^{+\infty}_{-\infty}
    (f_p-f_e)\,dv ~.
\end{eqnarray}
Here $a = e,p$, $m_{a}$ is the particle mass, $f_a$ the
component's DF, $\mu_{e}=1$, $\mu_{p}=-m_e/m_p$, and $E(x,t)$ is
the electric field. The boundary conditions are assumed to be
periodic. The initial electron distribution is
\[
f_e(x,v,0)=1/\sqrt{2\pi} v_{\tht e} \exp (- v^2/2 v^2_{\tht e}) [1
+ A_0 \cos(k_m x)]~,
\]
where $v_{\tht e}$ is the Maxwellian thermal speed for electrons,
$A_{0}$ the initial electric amplitude, $k_m=2\pi m/L$ is the wave
number of the mode $m$, and $L$ is the length of the system. The
ions are initially uniform and Maxwellian-distributed in velocity
space with $T_p=T_e$.

The simulations use $m=1$, $v_{\tht e}=0.4$, Debye length
$\lambda_{\mbox{\scriptsize D}e} \approx 0.31$, and $L=2 \pi \approx
20.18 \lambda_{\mbox{\scriptsize D}e}$. They have $N_x=256$ cells in
the $x$ direction both for electrons and ions, and $N_{ve}=20000$ and
$N_{vi}=2000$ cells in speed for electrons and ions, respectively,
within the domains $[-10 v_{\tht a},10 v_{\tht a}]$. The Cheng-Knorr
method \cite{che:kn} was used to solve Eqs (\ref{dst}) and
(\ref{pot}) with double precision. System invariants $I_{3a}=\int
f_a^3 dxdv$ are conserved better than $\Delta
I_{3e}/I_{3e}(0)<10^{-6}$ for electrons, and $\Delta
I_{3p}/I_{3p}(0)<10^{-9}$ for ions.

Fig.~\ref{f1} shows the evolution of the mode $m=1$ for initial
amplitude $A_0=0.012$, $A_{0}^{\star} = (8.51 \pm 0.06) \times
10^{-3}$, and $m_p/m_e=1836$. This type of evolution is observed
experimentally \cite{danielson}. The existence of significant ion
motion in Fig.~\ref{f1}(a) seems, at first glance, to suggest that
the evolution is seriously affected by ion mobility. However, the
envelope field amplitude of the electron oscillations in
Fig.~\ref{f1}(a) is almost identical to that for immobile ions
[Fig.~\ref{f1}(b)]. Quantitatively, the initial damping phase in
Fig.~\ref{f1} stops at time $t_{\tmn} \approx 441 \,
\omega^{-1}_{\pe}$ and amplitude $A_{\tmn} \approx 1.64 \times 10
^{-5}$, and is then followed by almost exponential growth which
saturates at $t_{\tst} \approx 1365 \, \omega^{-1}_{\pe}$ and
$A_{\tst} \approx 2.42 \times 10^{-4}$. These quantities are
identical to those calculated in the electron V-P simulations of
Ref. \cite{iv-ca-ro}, where $m=4$ was assumed for the perturbation
and $v_{\tht}=0.1$ for the electron thermal speed. This is
expected because $k\lambda_{\mbox{\scriptsize D}e}$, the wave
frequency $\omega$, and $\gamma_{\mbox{\scriptsize L}}$ are the
same for the two simulations.

Analytic theory predicts that $\omega \approx 1.2851 \,
\omega_{\pe}$, but the simulated value $\omega^\star = 1.2705 \pm
9 \times 10^{-4}$ is slightly shifted from $\omega$ due to the
large value of $A_{0}$ and varies slightly with time
[Fig.~\ref{f1}(b)]. Linear damping rate is
$\gamma_{\mbox{\scriptsize L}} \approx - 0.0661 \, \omega_{\pe}$.
For smaller $A_0=10^{-5}$ both $\omega^\star$ and
$\gamma_{\mbox{\scriptsize L}}$ match  the standard Landau theory
\cite{landau} very well (not shown), with $\{|(\omega ^\star -
\omega) / \omega |, |(\gamma ^\star_ {\mbox{\scriptsize L}} -
\gamma_{\mbox{\scriptsize L}}) / \gamma_{\mbox{\scriptsize L}}|\}
< 2 \times 10^{-4}$.

These two-component V-P results demonstrate that the threshold
phenomenon for Langmuir wave damping is robust against ion
effects. Accordingly one-component simulations, with ions acting
as a neutralizing background, are used below.

\begin{figure}[t]
\includegraphics[clip]{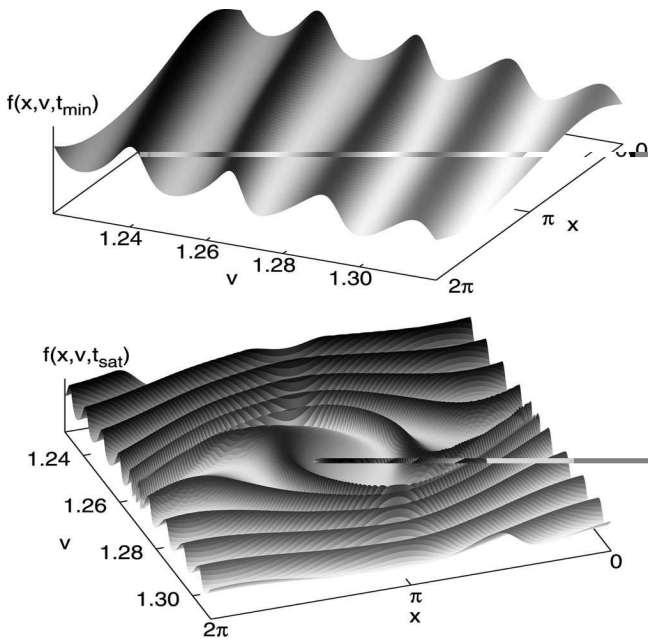}
\caption{DF when (upper panel) $t=t_{\tmn}$ and damping stops and
(bottom panel) $t=t_{\tst}$ and growth saturates.} \label{f2}
\end{figure}

The DF near the phase velocity $v_\phi = \omega^\star /k_1 \approx
1.271$ at these moments is shown in Fig.~\ref{f2} and reveals drastic
discrepancies between the evolution which ends with arrest of damping
at $t=t_{\tmn}$, and the subsequent evolution until the growth
saturates at $t=t_{\tst}$. At the moment $t=t_{\tmn}$ the phase space
portrait reveals no signs of particle trapping -- only filamentation
due to phase mixing (Fig.~\ref{f2}, the upper view). Moreover,
instead of a stationary state this distribution supports
approximately linear (meaning exponential) growth on the interval
$t_{\tmn} < t < t_{\tst}$, as Fig.~\ref{f1}(b) shows. Crucially, the
DF at $t_{\tst}$ does not consist of the closed orbits (or whorls in
velocity-position space) expected for trapping. Instead, the orbits
are still open, although they clearly indicate progress towards
trapping. Trapping is therefore responsible for neither the arrest of
damping nor the saturation of the growth phase.

\begin{figure}[b]
\includegraphics[clip]{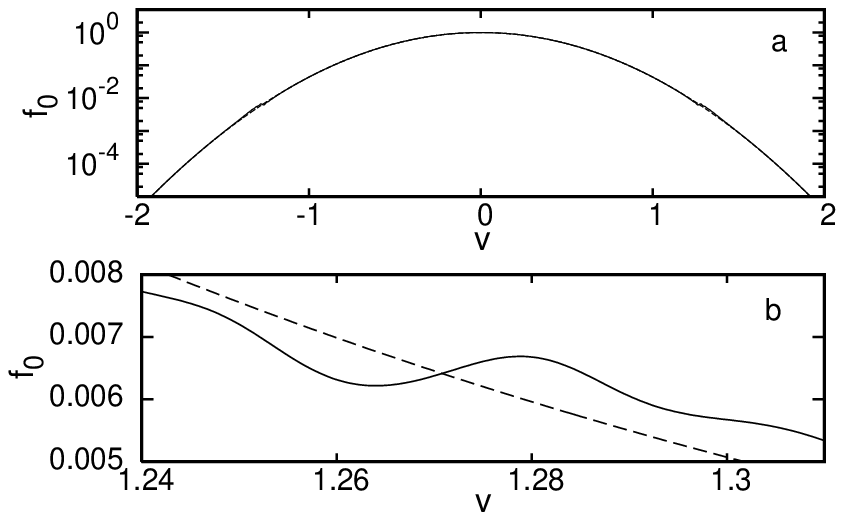}
\caption{$f_0(v,t_{\tmn})$ (solid line) and $f_0(v,0)$ (dashed line)
on semilogarithmic (upper panel) and linear (bottom panel) scales for
two velocity intervals: (a) $-2.0 \le v \le 2.0$ and (b) $1.24 \le v
\le 1.32$.} \label{f3}
\end{figure}

In the linear theory developed by Landau \cite{landau} growth is
due to a positive slope in the DF at the phase velocity of the
wave, $|v|=v_\phi$. Fig.~\ref{f3} shows the DF averaged on $x$
coordinate, $f_0(v,t) = (1/L)\int_0^L f(x,v,t)\,dx$ at
$t=t_{\tmn}$. Instead of the flattening of $f_0$ near the resonant
velocities $v=\pm v_\phi$ predicted by quasilinear theory
\cite{qlt}, $f_0(v,t_{\tmn})$ acquires a \textit{positive} slope
in a small vicinity of $v_\phi$, and therefore can support
(approximately) linear growth after the moment $t=t_{\tmn}$ as
Fig.~\ref{f1}(b) shows.

\begin{figure}[t]
\includegraphics[clip]{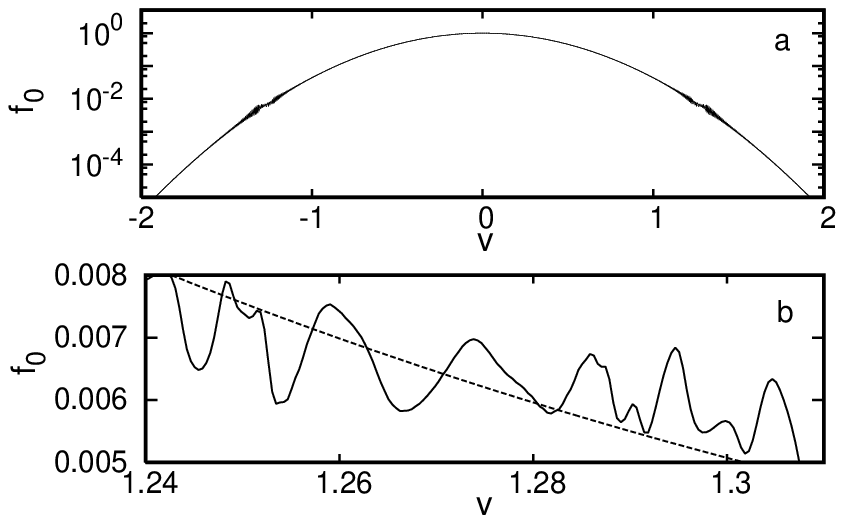}
\caption{$f_0(v,t_{\tst})$ (solid line) and $f_0(v,0)$ (dashed line)
on semilogarithmic (upper panel) and linear (bottom panel) scales for
two velocity intervals: (a) $-2.0 \le v \le 2.0$ and (b) $1.24 \le v
\le 1.32$.} \label{f4}
\end{figure}

Contrary to the situation near $t=t_{\tmn}$ when damping ceases
and the physics looks quite smooth and regular, $f_0$ becomes
quite irregular near the time $t=t_{\tst}$ when growth saturates
(Figs.~\ref{f2} and \ref{f4}). In particular, the lower panel of
Fig.~\ref{f2} is strongly reminiscent of trapping, although
strictly closed trajectories do not appear for this $A_{0}$. Also,
while on average the slope of $f_0(v,t_{\tst})$ at $v = \pm
v_\phi$ seems to have decreased compared with time $t=t_{\tmn}$
[Fig.~\ref{f4}(a)], it varies irregularly in the neighborhood of
$\pm v_\phi$ and therefore may support excitation of oscillations
with a wide range of phase speeds.

\begin{figure}[b]
\includegraphics[clip]{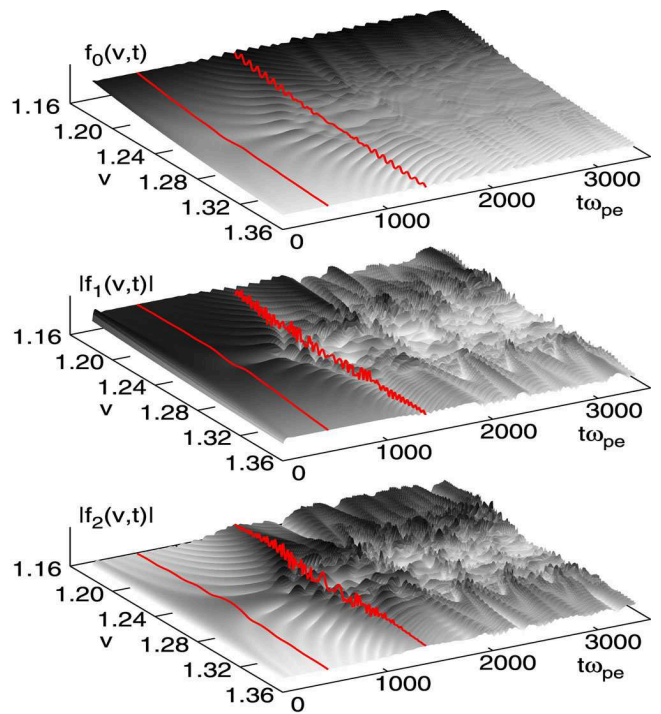}
\caption{Evolution of DF components $f_0$ (upper panel), $f_1$
(middle panel) and $f_2$ (bottom panel). Red lines are DF components
at the moments $t_{\tmn}$ and $t_{\tst}$, e.g. $|f_1(v,t_{\tst})|$.}
\label{f5}
\end{figure}

Fig.~{\ref{f5}} shows the evolution of, and power transfers between,
the average DF $f_0(v,t)$ and the DF components $f_1(v,t)$ and
$f_2(v,t)$ at $k_{1}$ and $k_{2}$, respectively, with $|f_m
(v,t)|=\{\mbox{Re}^2[f_m (v,t)] +\mbox{Im}^2[f_m (v,t)]\}^{1/2}$. It
shows that the dynamical picture can be divided into regions with
distinct characteristics that identify the processes causing the
evolution. Fig.~\ref{f5} shows that the turbulent processes
responsible for the (relative) flattening of $f_0$ in the resonant
area near $v_\phi$ start only \emph{after} $t_{\tst}$, when spatial
Fourier components $E_m$ other than $m=1$ become comparable to $E_1$
(not shown here).

\begin{figure}[t]
\includegraphics[clip]{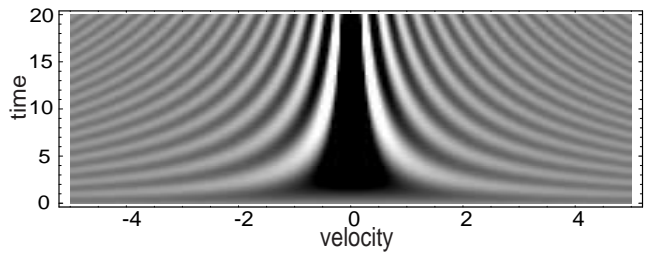}
\caption{Power transfer $P(v)$ for a damping wave with $\gamma=0.03$
in Eq. (15) of Ref.~\cite{drummond}).} \label{f6}
\end{figure}
\begin{figure}[b]
\includegraphics[clip]{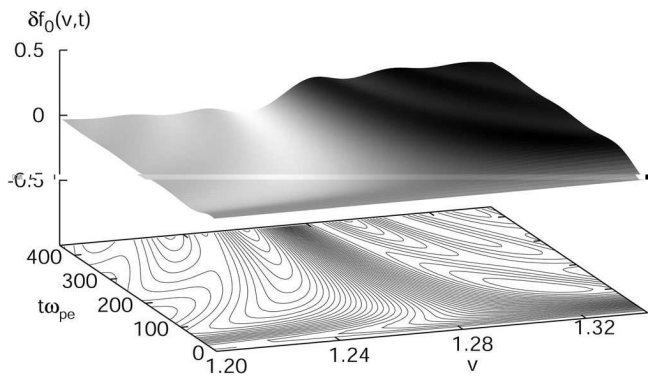}
\caption{Evolution of $\delta f_0(v,t)$ till the moment
$t=t_{\tmn}$.} \label{f7}
\end{figure}

The ripples of $f_0$, $|f_1|$, and $|f_2|$ in time and velocity
appear to be ``fingerprints'' of ballistic change of initial
perturbation and power transfer between the field and particles.
The latter claim is justified by Fig.~\ref{f6}, which illustrates
the power transfer rate for a wave growing/damping linearly by
resonant wave-particle interactions given by Eq. (15) in
Ref.~\cite{drummond}. Together with Fig.~\ref{f7}, which shows the
evolution of $\delta f_0(v,t)=[f_0(v,t) - f_0(v,0)]/f_0(v,0)$ on
the interval $0 \le t \le t_{\tmn}$, Figs~\ref{f5} and
Fig.~\ref{f6} clearly demonstrate that the physical process
responsible for arrest of linear damping is the resonant power
transfer between the wave and the $m=0$ and higher order
components of the DF.

An insight into the striking difference between the critical
exponents $\beta_{\tmn}$ and $\beta_{\tst}$ comes from critical
phenomena theory: critical exponents depend on the properties of
correlations for a specific system (e.g., on its dimensionality)
and/or a universality class (e.g., Ising, percolation, surface
growth etc.) \cite{goldn_hinr}. The DFs in full phase space
(position and velocity) are different at times $t_{\tmn}$
and $t_{\tst}$ (see Figs 2--5 and Fig.~\ref{f7}), so the critical
exponents might be different. This difference is contrary to the
idea that trapping explains both the arrest and saturation phases,
which should result in the \textit{same} exponents. Some
plausibility for velocity-space structures having this effect
follows from 1-D V-P
self-gravitating calculations: varying the resolution in $v$ seriously
affected estimates of the ``trapping scaling" exponent $\beta=2$
\cite{ivanov}.

In summary,  we studied the V-P model for initial Langmuir wave
amplitudes slightly above the threshold that separates damping and
non-damping evolution. Electron-ion simulations show that ion
mobility does not modify the threshold found for Langmuir damping
in electron-only simulations. Phase space diagnostics show no
signs of trapping or the DF flattening near $t=t_{\tmn}$ --
instead the combined effects of ballistic evolution of
perturbations and resonant power transfer at $|v| \approx v_\phi$
are responsible for arrest of the linear (Landau) damping then.
Since the spatially-averaged DF is not flat at $t_{\tmn}$ but
instead has a positive slope near the resonant velocity $v_\phi$,
this state is not stationary but instead leads to (linear) growth
which is saturated at $t=t_{\tst}$. The saturation time $t_{\tst}$
marks the boundary between the regular and stochastic evolution of
the wave electric field, again with no evidence for trapping
saturating the growth phase.


\begin{thebibliography}{99}

\bibitem{landau}
L. D. Landau, J. Phys. USSR \textbf{10}, 25 (1946).

\bibitem{sug-kam}
R. Sugihara and T. Kamimura, J. Phys. Soc. Jpn. \textbf{33}, 206
(1972); J. Canosa and J. Gazdag, Phys. Fluids \textbf{17}, 2030
(1974).

\bibitem{iv-ca-ro}
A. V. Ivanov, I. H. Cairns, and P. A. Robinson, Phys. Plasmas
\textbf{11}, 4649 (2004).

\bibitem{goldn_hinr}
N. Goldenfeld, \textit{Lectures on phase transitions and the
renormalization group} (Perseus Books, Reading, Mass.,
1992); H. Hinrichsen, Adv. Phys. \textbf{49}, 815 (2000).

\bibitem{ivanov}
A. V. Ivanov, \apj \textbf{550}, 622 (2001); A. V. Ivanov, S. V.
Vladimirov, and P. A. Robinson, \pre \textbf{71}, 056406 (2005).

\bibitem{strogatz}
S. H. Strogatz, R. E. Mirollo, and P. C. Matthews, \prl
\textbf{68}, 2730 (1992); S. H. Strogatz, Physica D \textbf{143},
1 (2000); J. A. Acebr\'on et al., \rmp \textbf{77}, 137 (2005); V.
Latora, A. Rapisarda, and S. Ruffo, Physica D \textbf{131}, 38
(1999).

\bibitem{crawford}
J. D. Crawford, \prl \textbf{73}, 656 (1994).

\bibitem{bump}
E. Frieman, S. Bodner, and P. Rutherford, Phys. Fluids \textbf{6},
1298 (1963).

\bibitem{qlt}
W. E. Drummond and D. Pines, Nucl. Fusion Suppl., \textbf{3}, 1049
(1962); A. A. Vedenov, E. P. Velikhov, and R. Z. Sagdeev, Nucl.
Fusion Suppl. \textbf{2}, 465 (1962).

\bibitem{oneil}
T. O'Neil, Phys. Fluids {\bf 8}, 2255 (1965); F. Valentini, V.
Carbone, P. Veltri, and A. Mangeney, \pre \textbf{71}, 017402
(2005).

\bibitem{la_do}
C. Lancellotti and J. J. Dorning, \prl \textbf{81}, 5137 (1998);
\pre \textbf{68}, 026406 (2003).

\bibitem{bgk}
I. B. Bernstein, J. M. Greene, and M. D. Kruskal, Phys. Rev.
\textbf{108}, 546 (1957); M. Buchanan and J. Dorning, \pre
\textbf{50}, 1465 (1994).

\bibitem{firpo}
M. C. Firpo and Y. Elskens, \prl \textbf{84}, 3318 (2000).

\bibitem{brun}
M. Brunetti, F. Califano, and F. Pegoraro, \pre \textbf{62}, 4109
(2000).

\bibitem{che:kn}
C. Z. Cheng and G. Knorr, J. Comput. Phys. \textbf{22}, 330
(1976).

\bibitem{danielson}
J. R. Danielson, F. Anderegg, and C. F. Driscoll, \prl
\textbf{92}, 245003 (2004).

\bibitem{drummond}
W. E. Drummond, Phys. Plasmas \textbf{11}, 552 (2004).

\end{thebibliography}
\end{document}